\documentclass[12pt,a4paper]{article}

\pdfoutput=1

\usepackage{exscale,picins,graphicx,hyperref}

\begin{document}

\title{Modelling an observer's branch of extremal consciousness} 

\author{L. Polley\\
        Institute of Physics\\ 
        Oldenburg University\\ 
        26111 Oldenburg, FRG}
\date{}

\maketitle

\begin{abstract}
Extreme-order statistics is applied to the branches of an observer in a many-worlds framework.
A unitary evolution operator for a step of time is constructed, 
generating pseudostochastic behaviour with a power-law distribution when applied repeatedly to a particular 
initial state. The operator models the generation of records, their dating, the splitting of the 
wavefunction at quantum events, and the recalling of records by the observer. 
Due to the huge ensemble near an observer's end, the
branch with the largest number of records recalled contains almost all ``conscious dimension''.   
\end{abstract}

\section{Introduction}

Extreme-order statistics, dealing with distributions of largest, second-largest values, etc., in
a random sample \cite{Embrechts2003}, has not received much attention in quantum theory. 
However, statistics of outliers can be striking in a many-worlds scenario,
due to the huge number of branches, providing the statistics is of the power-law type.  
If a random draw of some information-related quantity is made in each 
branch, the excess $l$ of the largest over the second-largest draw would be huge.
That excess exponentiates to $2^l$ if information is processed in qubits, 
each of which has two Hilbert-space dimensions. Thus, 
using the dimension as a weight of a branch \cite{Isham1994}, 
the weight of the extremal branch may exceed by far the total weight of all other branches. 
This might be a realisation, for an entire history of an observer 
rather than for a single measurement, of the idea that ``massive redundancy can cause certain 
information to become objective, at the expense of other information'' \cite{Zurek2005}.

The many-worlds scenario \cite{Everett1973} is implicit in decoherence theory 
\cite{Zurek1981} which has been able to explain why macroscopic superpositions 
evolve, rapidly, into states with the properties of classical statistical ensembles. 
There remains, however, the ``problem of outcomes'' \cite{SchlosshauerBuch2007}: 
Why is it that an observer of a superposition always finds himself ending up in a pure state  
instead of a mixture? 
The problem is not with objective facts, but with the consciousness of the observer. 
For, if any statement is derived from observations alone, it can only involve observations from one 
branch of the world because of vanishing matrix elements between different branches. 
Nowhere on the branching tree an objective contradiction arises.     
As to a conscious observer, however, we cannot tell by present knowledge whether he
needs to ``observe'' his various branches in order to become aware of them, or whether it 
suffices for him to ``be'' the branches, whatever that means in physical terms \cite{Penrose1997}.
In the model constructed below, the ``observer'' could in principle be aware of all of his branches,
but, due to statistics of extremes under a power law, this amounts to being aware of one extremal
branch plus fringe effects.  

We shall be modelling, rather abstractly, an observer ``as a finite in\-form\-a\-tion-processing structure'' 
\cite{Donald1999}, with an emphasis on ``processing'', because it means nothing 
to the observer if a bit of information ``has'' a certain value unless the value is revealed by interaction. 
We shall take a history-based approach mainly for reasons of extreme-value statistics under 
a power-law: outliers are most pronounced in large random samples, suggesting to use as a sample 
the entire branching tree. 
In addition, ``the history of a brain's functioning is an essential part of its nature as an object on 
which a mind supervenes'' \cite{Donald1997}. 

Ideally, a model of quantum measurement would be based on some Hamiltonian without explicit time 
dependence; in particular, without an external random process in time.
Such an approach via the Schr\"odinger equation, however, would be hindered by difficulties in solving the 
equation. We shall greatly simplify our task by considering time evolution only in discrete steps, and by
constructing a unitary operator directly for a time step. It would be possible, though of little use, 
to identify a Hamiltonian generating the evolution operator. 
A further simplification will be to consider evolution only during an observer's lifetime. 

If we describe measurements in the aforementioned way, 
we must show how stochastic behaviour can emerge \emph{in the observation}
of quantum systems. It will be sufficient to use random numbers in the construction of the time-step 
operator, in presumed approximation to real-world dynamical complexity. Under repeated application of the 
(constant) operator, evolution is deterministic in principle.
When the application is to a particular initial state, however, the built-in randomness becomes effective.

The evolution operator is constructed as a product of unitaries. This facilitates evaluations;
in particular, it enables a straightforward definition of ``conscious dimension''.   

The scenario of the model is as follows. A quantum system is composed of a record-generating part, like some 
kind of moving body; of records keeping track of the motion; and of subsystems associated with the records,
allowing for demolition-free reading. The observer appears through the subsystems and 
through part of the evolution operator. At his ``birth'', all records are in blank states, while the 
record-generating body is in some quasiclassical state which \emph{determines} the subsequent evolution. 
The evolution operator provides four kinds of event: Quasiclassical motion for most of the time, 
accompanied by the writing of records; dating of records by conservative ageing; splitting of the motion 
into a superposition of two equal-amplitude branches at certain points of the body's orbit; and the reading 
of records by a ``scattering'' interaction with the subsystems of the records. 

Random elements in the evolution are: Duration of quasiclassical sections;
the states of the body at which evolution continues after a split; and most crucially, the number of 
records being recalled within a timestep. For the latter number, a power-law
distribution is assumed\footnote{The standard mechanism for generating such a distribution is a 
supercritical chain reaction stopped by an event with a constant rate of incidence 
\cite{SimkinRoychowdhury2006}. Phenomenologically, power-law distributions are not uncommon in neurophysics,
but it seems they are always discussed in highly specialised contexts.}. No attempt is made here to 
justify the distribution---it should be regarded as a working hypothesis for the purpose of demonstrating 
the potential relevance of power-law statistics for quantum measurement.      
  
The state of superposition, emerging by repeated steps of evolution from an initial state of the chosen
variety, can be made explicit to a sufficient degree. It can be put in correspondence with a branching tree 
of the general statistical theory of branching processes. Statistical independence as required by that
theory is exactly satisfied by the model evolution, due to random draws employed in constructing the 
evolution operator.      
      
Consciousness is assumed to reside in unitary rotations of the recalling subsystems of the records, 
triggered when a record of a special class is encountered. The trigger is associated with a random draw 
determining the number of ``redundant'' records to be processed. Somewhere on the branching tree (that is, 
in some factor of the tensor products superposed) that number takes the extremal value. Because of the 
branching structure, the probability is large for that value to occur near the end of an observer's 
lifetime. Hence, it singles out (almost) an entire history. A study into the probability distribution
for the difference between the largest and the second-largest draw finally shows that the 
dimension of the subspace affected by conscious rotations in all branches of the superposition is
almost certainly exhausted by the dimension of conscious rotations in the extremal branch. 

The ``objective'' factors of the evolution operator are constructed in sections 
\ref{secQclStates} to \ref{secDating}. Their effect on the initial state 
is evaluated in section \ref{secTheSuperposition}. Conscious processing of records is modelled in 
section \ref{secRR}, while its statistics is analysed in section \ref{secExtremalBranch}.
Section \ref{Born} shows how the model would generalise to branching with non-equal 
amplitudes or into more than two branches. Conclusions are given in section \ref{Conclusion}.

\section{Construction of the evolution operator\label{secConstruction}}

\subsection{Record-generating system\label{secQclStates}}

A basic assumption of the model is that for all but a sparse subset of time steps, evolution is 
quasi-classical, like a moving body represented by a coherent state. Evolution on such a section is 
determined by a small set of dynamical variables, while a large number of ``redundant'' records is 
written along the path. In the model's approximation, instantaneous dynamical variables are represented 
by one number from the orbital set  
\begin{equation}  \label{TriggerIndicesSet}
    \{ 1,\ldots,K \} = : {\cal O} 
\end{equation}
The ordering is such that quasi-classical evolution takes the record-generating system from index $k$ to 
index $k+1$ within a step of time. The corresponding basis vectors of the record-generating system are
denoted by
\begin{equation}  \label{BasisRecordGeneratingSystem}
   \psi_k \qquad k \in {\cal O} 
\end{equation}
They are assumed to be orthonormal.

\begin{figure}
\caption{\label{RecordAgeing}States of a record represented by $N$ dots on a circle, 
ageing (without loss of information) under repeated writing operations.}
\center
\includegraphics[scale=1.0]{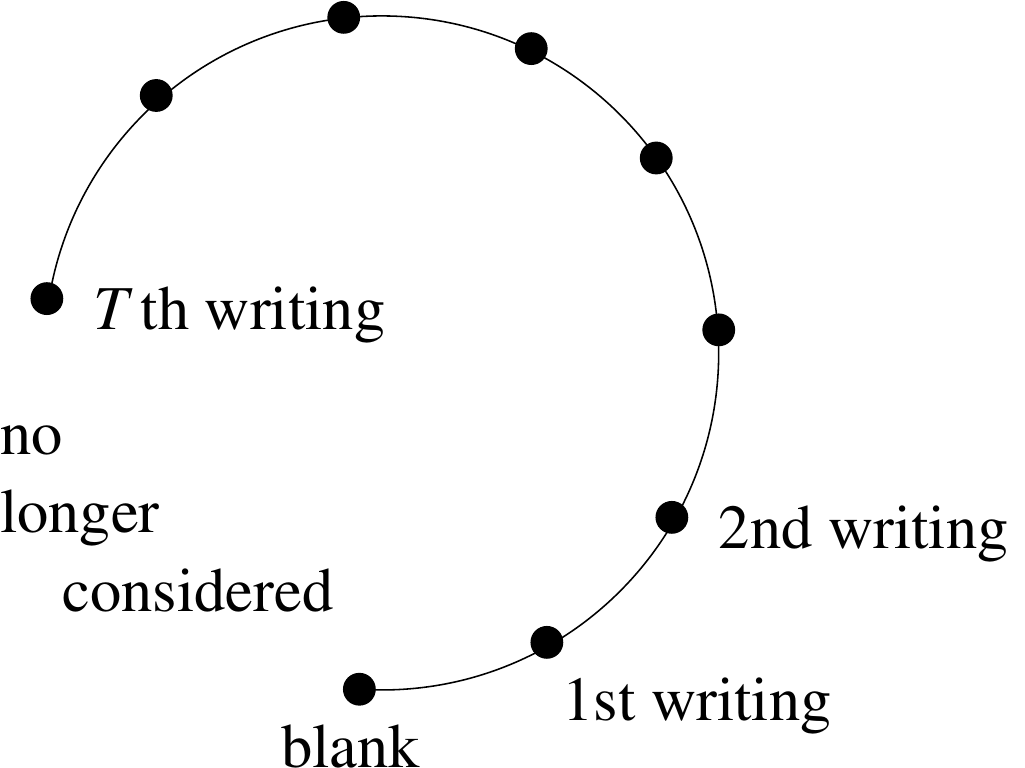}
\end{figure}

\subsection{Structure of records and the writing operation}

The focus of the model is on an observer's interaction with records generated during his personal 
history.  The term ``record'' will refer to 
observer's individual memories as well as to more objective forms of recording. Moreover, it will be 
convenient to use the term ``record'' as an abbreviation for ``recording unit''. Thus, records can be in 
``blank'' or ``written'' states. 

Each recording unit $r_i$ decomposes into a subsystem holding the 
information, and another subsystem allowing the observer to interactively read the information 
without destroying it. The information consists, firstly, of some quality represented by the index of the 
record, and secondly, in the time of recording, or rather the age of the record. The age is encoded in 
canonical basis vectors as follows. 
\begin{equation}  \label{AgeBasis}
   \begin{array}{l} e_0 = \mbox{blank state} \\
                    e_j = \mbox{written since $j$ steps of time} \qquad j = 1,\ldots,N-1
   \end{array}
\end{equation}
This is illustrated in figure \ref{RecordAgeing}.
The information about age will be crucial for composing an observer's conscious history by
one extremal reading. Only ages up to $N-1$ steps of time are possible, which is sufficient 
if we are dealing with a single observer. 

Both the generation of a record and its ageing can be described by a writing operation $W$. Acting on 
an indicated record, it acts on the first factor of (\ref{DefRecordState}) according to
\begin{equation}  \label{DefWrite} 
   \begin{array}{l}
   W e_i = e_{i+1} \qquad i=0,1,\ldots,N-1 \\
   W e_N = e_0
  \end{array}
\end{equation}
The second of these equalities is unwarranted, expressing an erasure of the record and thus 
limiting the model to less than $N$ steps of time, but there does not seem to be any better choice 
consistent with unitarity. Obviously, writing operations on different records commute,
\begin{equation}  \label{Wcommuting}
   W_i W_j = W_j W_i \qquad \mbox{for all }i,j
\end{equation}  

To allow for the observer's conscious interaction with a record, a two-dimensional factor space 
is provided, vaguely representing the firing and resting states of a neuron. Reading is modelled
as a scattering of some unit vector $s$ into some other unit vector $s'$, in a way that could, in an 
extended model, depend on the index and the age of the record. 
In the present model, only dimensions will be counted,
so no further specification of $s$ and $s'$ is required. 
The Hilbert space of a single record is thus spanned by product vectors of the form   
\begin{equation}  \label{DefRecordState}
    r_i = e_{n_i} \otimes s_i \qquad \mbox{with }~
  \left\{\begin{array}{l} \displaystyle e_{n_i}\mbox{ an $N$-dimensional canonical unit vector}\\
                         s_i \mbox{ a $2$-dimensional unit vector} \end{array}\right.
\end{equation}
The Hilbert space of all records possible is spanned by product vectors 
$$
   r_1 \otimes r_2 \otimes \cdots  \otimes r_I
$$
where $I$, in view of the redundancy required, is much larger than $K$. The set of all indices of
records will be denoted by $\cal I$.  It will be convenient to use the following abbreviation. 
\begin{equation}  \label{RecordStatesShorthand}
      V({\cal A}) ~ = ~ \parbox{60mm}{any tensor product of records in which all $r_i$
                                      with $i\in{\cal A}$ are blank}
\end{equation}

\subsection{Initial state\label{secInitialState}}

We assume that when the observer is ``born'' the record-generating system is in some quasiclassical
state $\psi_{k_\mathrm{in}}$ in which also the observer's identity is encoded. All records of the 
personal history are initially blank. Using abbreviation (\ref{RecordStatesShorthand}), the 
assumed initial state can be written as 
\begin{equation}  \label{InitialState}
      |\mathrm{in}\rangle = \psi_{k_\mathrm{in}} \otimes V({\cal I})
\end{equation}
This choice of an initial state will imply that, in the terms of \cite{Zurek2005},
we are restricting to a ``branching-state ensemble''. Such a restriction is necessary for 
pseudorandom behaviour to emerge under evolution by repeated action of a unitary time-step operator. 
By contrast, eigenstates of that operator would evolve without any randomness.

$k_\mathrm{in}$ is located on some string of quasiclassical events, as defined in section \ref{secRS}. 
It is this string, chosen out of many similar ones, that acts as a ``seed'' which determines the observer's 
pseudo-random history.

\subsection{Quasiclassical evolution and quantum events\label{secRS}}

\begin{figure}
\caption{\label{SplittingEvolution} (a) Sections of quasi-classical evolution (thick lines) which, when an 
index in the set $\cal Q$ is encountered, split into superpositions of continued quasi-classical evolution. 
Length of lines, and addresses after splitting, are random elements of the evolution operator.
When an initial state is chosen, defining an observer's origin and the ``seed'' for pseudo-random 
evolution, a branching tree results whose first and second branches are shown in (b).} 
\center
\includegraphics[scale=1.0]{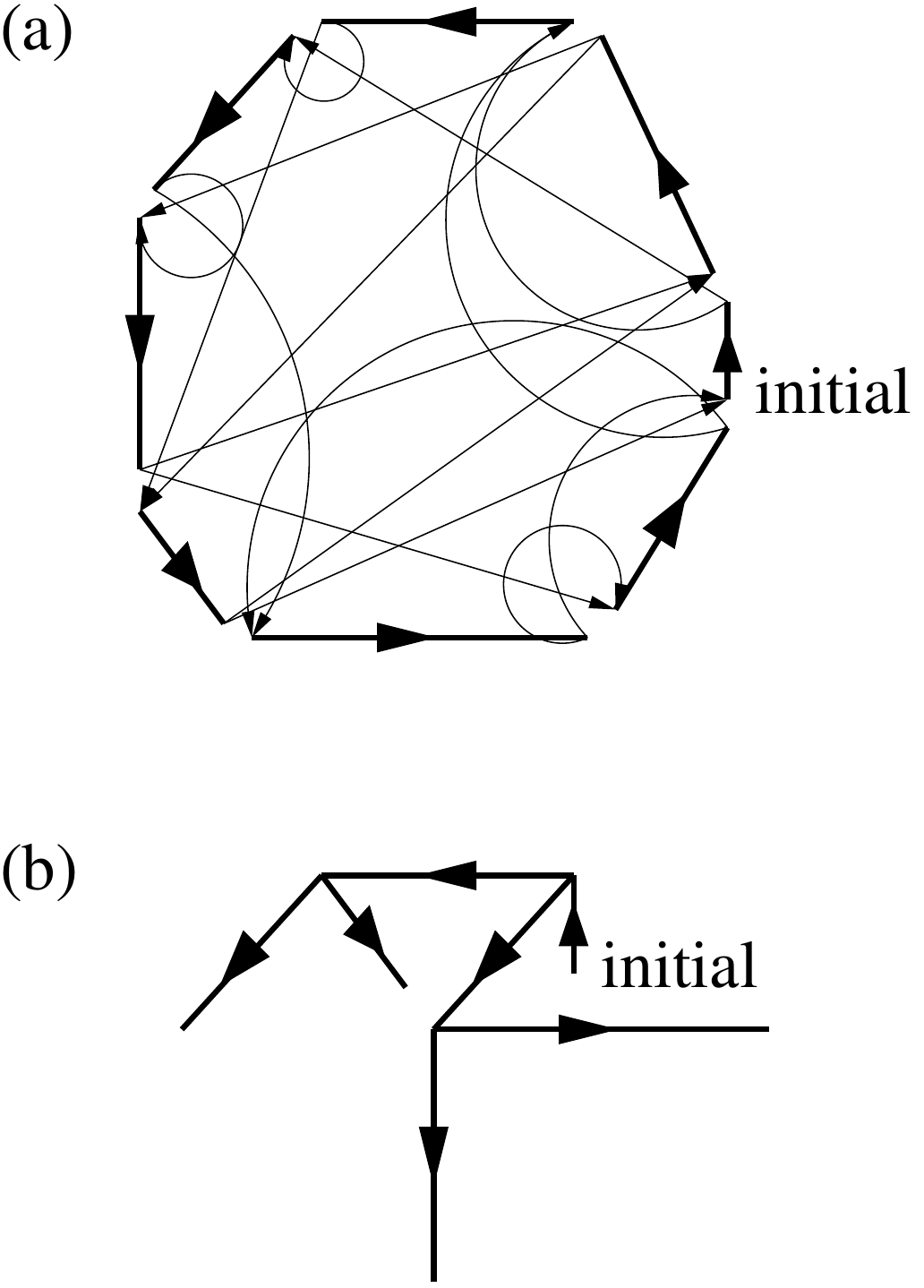}
\end{figure}

The idea of quasiclassical evolution, assumed to prevail for most of the time, is $\psi_k\to\psi_{k+1}$ in 
a timestep (section \ref{secQclStates}). This is to be accompanied by the writing of records. 
When the record-generating system is in the state $\psi_k$ we assume that  
writing operations $W_i$ act on all records $r_i$ whose indices, or addresses, are in a set 
${\cal A}_{\mathrm{W}}(k)$. While these records are redundant, we assume that $k$ can
be retrieved from each of them, which requires
\begin{equation}  \label{AW(k)intersection}
   {\cal A}_\mathrm{W}(k) \cap {\cal A}_\mathrm{W}(l) = \emptyset ~~\mbox{for}~~ k\neq l
\end{equation}  
When the record-generating system arrives at an index $k$ in a sparse subset 
${\cal Q}\subset{\cal O}$, we assume that a superposition of two branches (``up'' and ``down'') is formed, 
with equal amplitudes in both branches. Quasiclassical evolution is assumed to jump from $k$ to $u(k)$ or 
$d(k)$, respectively, and continue there, as illustrated in figure \ref{SplittingEvolution}. 

For simplicity of evaluation, we choose $u(k)$ and $d(k)$ to coincide with starting points of 
quasiclassical sections. These are indices in ${\cal O}$ subsequent to an index in ${\cal Q}$. 
We include index $1$ as a starting point and assume $K\in{\cal Q}$ to avoid ``boundary'' effects.  
Also, we must avoid temporal loops to occur within an observer's lifetime. That is, evolution operator 
$U_\mathrm{O}$ should be constructed such that no loops arise within $T$ repeated applications. 
Loops cannot be avoided entirely because 
for every $k\in {\cal Q}$ there are two jumping addresses, so every section of quasi-classical evolution 
will have to be targeted by two jumps on average.

To keep branches apart for a number $n$ of splittings, assume ${\cal Q}$ to be decomposable into $2^n$ 
subsets ${\cal J}[s]$, mutually disjoint, and each big enough to serve as an ensemble for a random draw. 
Let $s$ be a register of the form
\begin{equation}  \label{Register}
   s = [s_1,s_2,\ldots,s_n] ~~ \mbox{ where }~~ s_j \in \{ u,d \}
\end{equation}
Then, for $k\in{\cal J}[s_1,s_2,\ldots,s_n]$, define the jumping address $u(k)$ as follows. 
\begin{equation}  \label{Defu(k)}
   \begin{array}{l}
   \mbox{Draw $k'$ at random from } {\cal J}[s_2,s_3,\ldots,s_n,u]. \\
   \mbox{Put }u(k)\mbox{ at first point of quasiclassical section leading to }k'.
   \end{array} 
\end{equation}
Likewise for $v(k)$.
\begin{equation}  \label{Defv(k)}
   \begin{array}{l}
   \mbox{Draw $k'$ at random from } {\cal J}[s_2,s_3,\ldots,s_n,v]. \\
   \mbox{Put }v(k)\mbox{ at first point of quasiclassical section leading to }k'.
   \end{array} 
\end{equation}
The new entry, $u$ or $v$, will be in the register for $n$ subsequent splittings. Later on, 
the $u/v$ information is lost, allowing for inevitable loops to close. 
Given observer's lifetime $T$, and a minimal length $d_\mathrm{min}$ of quasiclassical sections, 
parameter $n$ should be chosen as $ n = T/d_\mathrm{min}$. The splitting addresses are collected in a set, 
\begin{equation}  \label{AkSuperposition}
 {\cal A}_\mathrm{S}(k) = \{u(k),d(k)\} 
\end{equation}  
Quasiclassical motion $k\to k+1$ and the branching $k\to(u,d)$ constitute the ``orbital''
factor of evolution which is to be described by a unitary operator $U_\mathrm{O}$. Under its action, 
images of orthonormal vectors must be orthonormal. Since loops are inevitable in the set of indices,   
orthogonality of images cannot be ensured by the states of the record-generating system alone, but can be
accomplished by orthogonalities (blank vs.\ written) in the accompanying states of records\footnote{There 
is a rudiment of decoherence in this model.}. To this effect, certain records must be blank before the 
action of $U_\mathrm{O}$. Their addresses are  
\begin{equation}  \label{UQBlankRecords} 
  {\cal B}(k) = \mbox{union of all }{\cal A}_\mathrm{W}(l)\mbox{ with nonempty }
    {\cal A}_\mathrm{S}(l) \cap {\cal A}_\mathrm{S}(k) \qquad k\in{\cal Q} 
\end{equation}
Using this, definitions (\ref{Defu(k)}), (\ref{Defv(k)}), and abbreviation (\ref{RecordStatesShorthand}), 
we define
\begin{equation}  \label{DefUO} 
    \begin{array}{ll}
    \displaystyle
    U_\mathrm{O}~\psi_k \otimes V({\cal B}(k)) = 
    \frac{ \psi_{u(k)} + \psi_{d(k)} }{\sqrt2} 
    \otimes \left(\prod_{i\in{\cal A}_\mathrm{W}(k)} W_i\right) V({\cal B}(k)) 
    & \quad k\in{\cal Q} \\ 
   \displaystyle 
   U_\mathrm{O}~\psi_k \otimes V({\cal A}_{\mathrm{W}}(k)) =
    \psi_{k+1} \otimes 
  \left(\prod_{i\in{\cal A}_\mathrm{W}(k)} W_i\right) V({\cal A}_{\mathrm{W}}(k)) & \quad k\notin{\cal Q} 
  \end{array}
\end{equation} 
This is a mapping of orthonormal vectors onto orthonormal image vectors, so it can be extended to a 
definition of a unitary operator $U_\mathrm{O}$ by choosing any unitary mapping between the orthogonal 
complements of the originals and images. However, due to the loop-avoiding construction, only 
properties (\ref{DefUO}) are required for the evolution of the initial states of section 
\ref{secInitialState}.

\subsection{Dating of records\label{secDating}}

The time at which a record was written can be retrieved from information about its age. 
An ageing operator for the $i$th record, $U_\mathrm{A}(i)$, is defined by its action on the first
tensorial factor of (\ref{DefRecordState}) as follows.    
\begin{equation}  \label{DefUA(i)} 
   \begin{array}{l}
   U_\mathrm{A}(i) e_0 = e_0 \\
   U_\mathrm{A}(i) e_i = e_{i+1} \qquad i=1,\ldots,N-2 \\
   U_\mathrm{A}(i) e_{N-1} = e_1 
   \end{array}
\end{equation} 
In particular, a record in blank state $e_0$ remains unchanged. Also, tensorial factors with indices 
different from $i$ are unaffected by $U_\mathrm{A}(i)$. At the limiting age, corresponding to $N$ 
steps of time, $U_\mathrm{A}(i)$ becomes senseless. 
The ageing operator for the entire system of recording units is 
\begin{equation}  \label{DefUA}
   U_\mathrm{A} = \prod_{i=1}^I U_\mathrm{A}(i)
\end{equation}  
The ageing of records is conservative, without loss of information.

\subsection{Explicit form of superposition \label{secTheSuperposition}}

The operators constructed in (\ref{DefUO}) and (\ref{DefUA}) define the ``objective'' 
part of time evolution. They do not act on the second tensorial factor of a record 
(equation (\ref{DefRecordState})). They are assumed to multiply in the order
\begin{equation}  \label{Uobj}
             U_\mathrm{O} U_\mathrm{A} = U_\mathrm{obj}
\end{equation}
Starting from our preferred initial product state $|\mathrm{in}\rangle$ (section \ref{secInitialState}), 
let us formulate the sequences $\{b_n\}$ of indices to which evolution branches under the action of 
$U_\mathrm{obj}$.   
If quasiclassical evolution has promoted the record-generating system to an index $k\in{\cal Q}$, 
branching to addresses $j\in{\cal A}_\mathrm{S}(k)$ occurs in the next step of time. 
A choice of $j$, here renamed $b_n$, characterises the branch. From $b_n$, quasiclassical evolution 
proceeds through indices numerically increasing until 
the next index in ${\cal Q}$ is reached, and branching to $b_{n+1}$ occurs. This takes a number of steps, 
$d_n$. 
The possible sequences of branching addresses $b_n$ and intervals $d_n$ of quasiclassical evolution 
must satisfy the following recursion relations.  
\begin{equation}  \label{BranchingSequence}
   \begin{array}{rcl}
   b_0 &=& k_\mathrm{in} \\
   q_n & = & \min \big\{ q \in {\cal Q} ~ | ~ q > b_n \big\} \quad 
   \mbox{(auxiliary)} \\
   b_{n+1} & \in &  {\cal A}_\mathrm{S}(q_n) \\
   d_n &=& q_n - b_n + 1 
   \end{array}
\end{equation}
The time $t_n(b)$ at which a branching index $b_n$ is reached, depending on the branch considered, is
\begin{equation}  \label{DefTau}
   t_n(b)  = \sum_{m=0}^{n-1} d_m(b)
\end{equation}
For a convenient representation of the stages of evolution in various branches, let us use
the following abbreviation.  
\begin{equation}  \label{integerTheta}
   [t] = \left\{\begin{array}{ll} 0 & \quad t \leq 0 \\
         t & \quad t > 0 
         \end{array} \right\} = \Theta(t-\epsilon)
\end{equation} 
Moreover, let $p(b,t)$ be the number of branching points passed by time $t$.
Referring to definitions (\ref{BranchingSequence}) and (\ref{DefTau}),  
the record-generating system then is in the state 
\begin{equation}  \label{psi(b,t)}
  \psi_{k(b,t)} ~~ \mbox{with}~~ k(b,t) = b_{p(b,t)} + t - t_{p(b,t)}
\end{equation}
Denoting by $V({\cal I})$ the state in which all records are blank, 
the evolved state after $t$ steps of time may be expressed as
\begin{equation}  \label{UrwsPsi}
   \left( U_\mathrm{obj}\right)^t |\mathrm{in}\rangle = \sum_b  \left(\frac1{\sqrt2}\right)^{p(b,t)} 
  \psi_{k(b,t)} \otimes
  \left( \prod_{n=0}^\infty \prod_{l=0}^{d_n-1}
   ~ \prod_{i\in{\cal A}_{\mathrm{W}}(b_n+l)} W_i^{[t - t_n - l]} \right) V({\cal I})
\end{equation} 
To see this, first note that once a record is written, its ageing is the same as repeated 
writing by (\ref{DefWrite}) and (\ref{DefUA(i)}). 
Writing operations can be assembled to powers because they commute (equation (\ref{Wcommuting})). 
Thus, the linear rise of the powers with $t$ is the result of the ageing operator $U_\mathrm{A}$.
It remains to consider the time of the first writing of a record. 
At time $t_n+l$, the record-generating system is in the state with index $b_n+l$. Corresponding 
records, with indices in ${\cal A}_{\mathrm{W}}(b_n+l)$, are written at the next step of time,
that is, when the exponent $[t - t_n - l]$ of the writing operator is nonzero for the first time.

For later reference we note that the product vectors constituting different branches are orthogonal. 
This is because two branches differ by at least one record, so that there is at least one tensorial 
factor $r_i$ which is in the blank state in one branch and written, hence orthogonal, in the other.

\subsection{Consciousness modelled as triggered recall\label{secRR}}

A third factor of the evolution operator is supposed to model reflections in the observer's mind. 
Neurophysical detail is beyond the scope of this paper, but consciousness ``supervenes'' on neural 
\emph{dynamics} \cite{Donald1999}.
The other ingredient of the present model is power-law statistics, which appears 
to be common in neurophysics, but is usually discussed in highly specialised context. 
It is essentially a working hypothesis here.    

As this part of the evolution operator is going to make its dominant impact near the end of an
observer's histories, it must be prevented from writing records of an objective sort, that is, 
from writing any records at all in the model's terms. 
Otherwise, a scenario would result in which the facts constituting an observer's histories
would be generated within a step of time. Thus the factor $U_\mathrm{C}$, defined below, should be 
only reflective, like reading a record by elastic scattering.  Activities like  
writing this article would be regarded as ``subconscious'', that is, rather a matter of 
$U_\mathrm{O} U_\mathrm{A}$ within the scope of the model.  

\subsubsection{Triggering records}

Conscious reflection is assumed to be triggered by the reading of a record $r_m$ in a sparse 
index set ${\cal M} \subset {\cal I}$. Moreover, it is assumed for simplicity that 
\begin{equation}  \label{OneMemoryindexAW} 
 \mbox{for each }k\in{\cal O}\backslash{\cal Q} 
         \mbox{ there is exactly one such $m$ in } {\cal A}_\mathrm{W}(k)  
\end{equation} 
If $r_m$ is blank, no reflection occurs. If $r_m$ is in a written state,
a ``scattering'' operation $S_l$ will be triggered on all $r_l$ with indices in a set 
${\cal A}_{\mathrm{R}}(m)$ specified in equation (\ref{DefAR}) below. 

Let $P^0_m$ denote the projection on the blank state of $r_m$, and $P^\perp_m$ 
the projection on all written states of $r_m$. We define the reflection triggered by $r_m$ as
\begin{equation}  \label{DefUrr} 
   U_\mathrm{C}(m) = P^0_m + P^\perp_m \prod_{l\in{\cal A}_{\mathrm{R}}(m)} S_l 
\qquad m \notin{\cal A}_{\mathrm{R}}(m) 
\end{equation} 
with implicit unities for all tensor factors whose indices do not appear.
A scattering operation $S$, in the indicated space, is assumed to modify the
second factor of (\ref{DefRecordState}) in a way dependent on the first factor, 
\begin{equation}  \label{DefRecall} 
   \begin{array}{l}
   S_l ~ e_0 \otimes s  = e_0 \otimes s \\
   S_l ~ e_i \otimes s  = e_{i} \otimes u_{li} s \qquad u_{li} \neq 1 \qquad i=1,\ldots,N-1
   \end{array}
\end{equation} 
For records in a written state, all we assume about the unitary $2\times 2$ matrices 
$u_{li}$ is that they be different from $\bf 1$ so as to make ``something'' go on in the observer's 
mind.
 
A crucial assumption is made on the statistics, in random draws for the construction, 
of the lengths $L({\cal A}_{\mathrm{R}})$ of the address sets ${\cal A}_{\mathrm{R}}$. 
Let $\overline{F}(L)$ be the complementary cumulative distribution function, that is the fraction of sets 
whose length is greater than $L$. We assume a capped Pareto distribution  
\begin{equation}  \label{AkPowerLaw}
            \overline{F}_1(L) = \left\{\begin{array}{cr} (L_0/L)^\alpha &  L_0 \leq L \leq I \\
       0 & L > I \end{array} \right\} \qquad 1 < \alpha < 2 
\end{equation}
where the cap is assumed to be practically irrelevant due to the size of the index set $\cal I$. 
To ensure the statistical independence required for the theorems of order statistics to apply, 
let us construct the index sets by explicit use of independent random draws. 
In a first step, the lengths of sets are determined. 
\begin{equation}  \label{RandomDrawSize} 
 \mbox{For all $m\in{\cal M}$,} ~ L(m) = \mbox{random draw from distribution (\ref{AkPowerLaw})} 
\end{equation}
In a second step, $L(m)$ indices are selected by another random procedure, and collected into
${\cal A}_{\mathrm{R}}(m)$. The procedure is as follows.

\subsubsection{Searching for potential records \label{RetrieveRecords}}

Operator $U_\mathrm{C}$ must select $L(m)$ records that \emph{may} have been written during time evolution.
In fact, if recall operations were searching for records irrespective of causal relations, the scenario 
envisioned would not work statistically. The search would be based on mere chance---on a \emph{probability} 
proportional to $L(m)$, which would either have to be very small, or could not be power-law distributed, 
since probabilities are bounded above.  

Tracing back histories that may have lead to a memory index $m$, there emerges a backward-branching 
structure because there are, on average, two indices of ${\cal Q}$ from which quantum jumps are directed 
to a given section of quasiclassical evolution; see section \ref{secRS}.  
Starting from the memory-triggering index $m$, all sequences $\{c^m_n\}_{n=0,1,\ldots}$ of branching points 
that may have lead to the writing at $m$ must satisfy the following relations. 
\begin{equation}  \label{BackwardBranchingSequence}
   \begin{array}{rcl}
   c^m_0 &=& \{ k\in{\cal O} ~|~ m \in {\cal A}_\mathrm{W}(k)\} \\
   j_n & = & 1 + \max \big\{ q \in {\cal Q} ~ | ~ q < c^m_n \big\} \quad 
  \mbox{(auxiliary\footnotemark)} \\
   d^m_n &=& c^m_n - j_n  \quad \mbox{(length of quasiclassical section)} \\
   j_n & = & {\cal A}_\mathrm{S}(c^m_{n+1}) \quad \mbox{(preceding points of branching)} 
   \end{array}
\end{equation}
\footnotetext{Tracing back quasiclassical evolution, which is index-increasing, $j_n$ is the first index 
              encountered to which evolution may have jumped from somewhere.}
Indices $c^m_n,c^m_n-1,\ldots,c^m_n-d^m_n$ constitute the $n$th section on a branch of \emph{possible} 
evolution. The average length of such a section is $K/Q$.
We wish to distribute $L(m)$ conscious recalls equally over a lifetime. Hence there are
\begin{eqnarray}  
               L(m)/T && \mbox{ recalls per time} \label{RecallsPerTime} \\
   l(m) =  K L(m)/ TQ && \mbox{ recalls per section} \label{effectiveL}
\end{eqnarray}
Thus, for every sequence $c$ branching backward from $m$ and for every section number $n$ let us define
$$
    \begin{array}{rl} 
    {\cal C}(m,c,n) ~ = & \mbox{set of $l(m)$ randomly chosen elements unequal $m$} \\ & \mbox{of } 
   {\cal A}_\mathrm{W}(c^m_n) \cup{\cal A}_\mathrm{W}(c^m_n-1) \cup \cdots 
    \cup {\cal A}_\mathrm{W}(c^m_n-d^m_n)
    \end{array}
$$
In terms of ${\cal C}(m,c,n)$ we can specify the index sets already used in (\ref{DefUrr}).
\begin{equation}  \label{DefAR}
   {\cal A}_\mathrm{R}(m) = \bigcup_{n=0}^{TQ/K} ~ \bigcup_{\{c\}} ~ {\cal C}(m,c,n) 
\end{equation}
The full consciousness-generating part of the evolution operator is, referring to (\ref{DefUrr}) again, 
\begin{equation}  \label{Urecall}
   U_\mathrm{C} = \prod\limits_{m\in{\cal M}} U_\mathrm{C}(m)
\end{equation}

\section{Branch of extremal consciousness\label{secExtremalBranch}}

By $T$ steps of evolution, a superposition of product states builds up, which in equation 
(\ref{UrwsPsi}) was expressed as a sum over branches, each branch being generated by a product of writing 
operations. One-to-one correspondence to a branching tree can be seen by factoring out $W$ operations
of common parts of the branches. The loop-avoiding construction of section \ref{secRS} is important here. 

On the branching tree, certain memory-triggering records are in a written state. One 
of those records will trigger the maximal number of recalls, whose excess we wish to quantify 
statistically. It would be straightforward to estimate the excess on the basis of mean values alone,
similar to the argument given in \cite{Polley2008}, 
but fluctuations in branching processes are as big as the mean values \cite{Harris1963} so analysis 
in terms of probability distributions is required.

\subsection{Statistics of branching and recall-triggering}

The general theory of Galton-Watson processes \cite{Harris1963} deals with 
familiy trees whose members are grouped in generations $n=1,2,3,\ldots$. Each member generates 
a number $j=0,1,\ldots$ of members of the next generation with probability $p_j$.
 In our model, a new generation occurs at each step of time. The number of members in a generation,
$Z_t$, is the number of product states superposed at time $t$. 
The probability $p_0$, corresponding to an end of a branch of the observer's history, is zero within the 
lifetime $T$ considered. 
The probability $p_1$, corresponding to a product state continuing as a product state 
after a step of time, is close to one. The probability $p_2$, corresponding to the splitting
of a branch into a superposition of two product states, is small but nonzero. 
Probabilities $p_3,p_4,\ldots$ are zero by the model assumptions. 

In our model, splitting in two branches occurs at $Q$ randomly distributed points of $K$, so the parameters 
for the branching process here are
\begin{equation}   \label{BranchingParameters} 
   p_2 = \frac{Q}{K} = : \sigma  ~~~~~~~~~ p_1 = 1 - p_2 ~~~~~~~~~~ 
   p_j = 0 \mbox{ for }j = 0,3,4,5,\ldots
\end{equation} 
The mean number of offspring generated by a member thus is
\begin{equation}   \label{DefBranchingMeanValue} 
   \mu = 1 + \sigma > 1
\end{equation}  
Because of $p_0=0$, we are dealing with zero ``probability of extinction''.

For the statistics of the extremes, we need to know the total number of recall-triggering factors on the 
tree. By assumption (\ref{OneMemoryindexAW}) that number equals the ``total progeny'' 
$Y_t = \sum_{\tau=1}^t Z_\tau$.
By Theorem 6 of \cite{Pakes1971}, the probability distribution for the values of $Y_t$ has an asymptotic 
form which can be described as follows. There exists a sequence of positive constants 
$C_t$, $t=1,2,\ldots$, with $C_{t+1}/C_t \to \mu $ for $t\to\infty$ such that     
\begin{equation}   \label{YtProbability}
   \lim_{t\to\infty} P\{Y_t \leq x C_t\} = P\{W\leq x\sigma/\mu\}
\end{equation} 
where $W$ is a non-degenerate random variable which has a continuous distribution on the set of positive
real numbers. Let $w$ be the probability density for $W$. We shall treat $Y$ as continuous, too, and
assume that by an observer's lifetime $T$ the limiting form of (\ref{YtProbability}) already applies. 
Then the probability of $Y$ is, by differentiating (\ref{YtProbability}) and using $\mu\approx 1$,
\begin{equation}   \label{DensityMemorisingIndices}
          w\left(\frac{\sigma Y}{C_T}\right)\,\frac{\sigma}{C_T} \, \mathrm{d}Y
\end{equation}
Each occurrence of a memory-triggering index $m$ is characterised by the location on the tree, 
in particular the time $t$, and by the length $L(m)$ of the recalling sequence according to 
(\ref{RandomDrawSize}). Since the location results from random draws in $U_\mathrm{O}$, 
and the length from a random draw in $U_\mathrm{C}$, they are statistically independent, so their joint 
probability is the product of the separate probabilities. 
The time $t$ of occurrence, that is the generation number in the general theory, has a probability 
$Z_t/Y_t$ whose asymptotic form, under the same conditions as for (\ref{YtProbability}), is given by 
Lemma 2.2 of \cite{Pakes1998}. If $j = 0,1,2,\ldots$ denotes the distance form the latest time on the tree, 
the probability is
$$
    P_j = (1-\mu^{-1}) \mu^{-j}
$$
Taking the latest time on the branching tree to be $T$, and treating $L$ as continuous, 
$P_j$ and the Pareto distribution (\ref{AkPowerLaw}) give the joint probability of $t$ and $L$, 
\begin{equation}   \label{JointProbabilityLt}
   P(t,L)\, \mathrm{d}L = P_{T-t} \,\alpha L_0^\alpha L^{-\alpha - 1} \, \mathrm{d}L 
  \qquad 0\leq t \leq T, ~ L \geq L_0  
\end{equation}
If the memory-triggering $m$ occurs at $t$, then by (\ref{RecallsPerTime}) the number of records recalled is
\begin{equation}   \label{defR}
    R = \frac{t}{T}\,L(m)
\end{equation} 
It is the extreme-order statistics of this quantity that matters. The density of $R$ is obtained by
taking the expectation of $\delta(R - tL/T)$ with the probability distribution (\ref{JointProbabilityLt}).
In the range $R>L_0$ this gives another Pareto distribution with complementary cumulative distribution 
function
\begin{equation}   \label{DensityIndicesRecalled}
    \overline{F}_2(R) = (R_0/R)^\alpha ~ \mbox{ where } ~ R_0^\alpha = 
   L_0^\alpha \sum_{t=0}^T P_{T-t} \left(\frac{t}{T}\right)^\alpha
\end{equation} 
Thus, with probability 
given by (\ref{DensityMemorisingIndices}), we have a number $Y$ of memory-triggering indices on the 
branching tree of a lifetime, each of which with \nolinebreak a probability given by 
(\ref{DensityIndicesRecalled}) induces $R$ recalls along its branch. 

We now use a result of order statistics, conveniently formulated for our purposes in \cite{Embrechts2003},
table 3.4.2 and corollary 4.2.13, which relates the number of random draws, 
here $Y$ (different letters used in \cite{Embrechts2003}), to the spacing $D$ between the largest and the 
second-largest draw of $R$ from an ensemble given by (\ref{DensityIndicesRecalled}). 
\begin{equation}   \label{FrechetSeparation}
    D = R_{\mbox{\small largest}} - R_{\mbox{\small second-largest}} = R_0 \, Y^{1/\alpha} \, X 
\end{equation}
where $X$ is a random variable independent of $Y$. The probability density $g(x)$ of $X$, given in 
integral representation, can be seen to be uniformly bounded. 
The cumulative distribution function for $D$ is, for a given value of $Y$, of the form 
$G\left(Y^{-1/\alpha}D/R_0\right)$ where $G'(x)=g(x)$. 
Hence, the joint probability of $Y$ and $D$, expressed by density (\ref{DensityMemorisingIndices}) for $Y$ 
and the cumulative distribution function for $D$, is 
\begin{equation}   \label{JointProbabilityYD}
  G\left(Y^{-1/\alpha}D/R_0\right)
  w\left(\frac{\sigma Y}{C_T}\right)\,\frac{\sigma}{C_T}\,\mathrm{d}Y  
\end{equation}

\subsection{Dimension of conscious subspace\label{secConsciousDimension}}

Consciousness, in the model's approximation, is assumed to reside in unitary rotations $u\neq 1$ of the 
right tensor factors of (\ref{DefRecall}). Transformations of the left factors, 
as generated by $U_\mathrm{obj}$, are assumed to be unconscious. 
In the superposition generated by $U_\mathrm{obj}$, equation (\ref{UrwsPsi}),
the branches (product vectors) are mutually orthogonal by the ``objective'' left factors alone, 
as was noted at the end of section \ref{secTheSuperposition}. Hence, the unitary rotations of 
consciousness take place in subspaces which, for different branches, are orthogonal. 
Thus, a ``conscious dimension'' $d_\mathrm{C}$ can be assigned to each branch.
\begin{equation}  \label{DefConsciousDimension}
   d_\mathrm{C} ~ = ~ \parbox[t]{100mm}{Hilbert-space dimension of the tensor factors rotating under 
                                $U_\mathrm{C}$ while the remainder of factors is constant.}
\end{equation}
The number of tensor factors rotating in a branch is $R$, as defined in equation (\ref{defR}), so the 
dimension of the conscious subspace in a branch is $2^R$. It should be noted that the subspace
as such \emph{depends} on the vectorial value taken by the nonrotating factors. 

The proposition of the paper is that the conscious dimension in the branch with the largest $R$
exceeds, by a huge factor $E$, the sum of conscious dimensions in all other branches. The latter sum can
be estimated, denoting by $Z_T$ the number of branches (terms of superposition) at time $T$, as 
$$
    < ~ 2^{R_{\mbox{\small second-largest}}} Z_T
$$
Evaluating this would require a joint distribution of $R$, $Y$, and $Z$, so a more convenient estimate,
using $Z_T < Y_T$, is 
$$
    < ~ 2^{R_{\mbox{\small second-largest}}} Y_T
$$
Taking binary logarithms, the customised proposition is that the last term in the equation
\begin{equation}   \label{PropositionCustomised}
     R_{\mbox{\small largest}} =  R_{\mbox{\small second-largest}} + \log_2 Y  + \log_2 E 
\end{equation} 
almost certainly takes a large value. By (\ref{FrechetSeparation}), $\log_2 E = D - \log_2 Y$, 
so the relevant cumulative distribution function is obtained from (\ref{JointProbabilityYD}) as
\begin{equation}   \label{defF3}
   F_3(x) = P\{D - \log_2 Y < x \} = \int_1^\infty \!\! G\Big(Y^{-1/\alpha}(x + \log_2 Y)/R_0\Big) 
       w\left(\frac{\sigma Y}{C_T}\right)\,\frac{\sigma}{C_T} \,\mathrm{d}Y
\end{equation}
Substituting
\begin{equation}   \label{DifferenceRescaled}
     \frac{\sigma Y}{C_T} = y ~~~~~~~~~~~~ \left(\frac{\sigma}{C_T}\right)^{1/\alpha} \frac{x}{R_0} =  z 
\end{equation}
and putting $\sigma/C_T\approx 0$ in the lower limit of integration, the integral becomes
$$
   \int_0^\infty G\left(y^{-1/\alpha} z + \left(\frac{\sigma}{C_T}\right)^{1/\alpha} R_0^{-1}
   y^{-1/\alpha}(\log_2 y + \log_2 C_T - \log_2\sigma)\right) 
       w(y) \,\mathrm{d}y
$$
Asymptotically for $C_T\to\infty$, which represents an exponentially grown number of branches, 
the integral simplifies to 
$$
   F_3(x) = \int_0^\infty G\left(y^{-1/\alpha} z )\right) w(y) \,\mathrm{d}y
$$
because $G$ has the uniformly bounded derivative $g$ (see text following (\ref{FrechetSeparation})) while 
$\int_0^\infty y^{-1/\alpha} \log_2 y \, w(y) \,\mathrm{d}y$ converges for $\alpha$ in the range 
given by (\ref{AkPowerLaw}) and the coefficients $C_T^{-1}$ and $C_T^{-1}\log_2 C_T$ become vanishingly 
small. Inserting $z$ from (\ref{DifferenceRescaled}), and $x=\log_2 E$ from (\ref{defF3}),  
the cumulative distribution function for the excess factor $E$ is given by
\begin{equation}   \label{cdfH}
    F_3\left(\left(\frac{\sigma}{C_T}\right)^{1/\alpha}R_0^{-1} \log_2 E\right) 
\end{equation}
Due to logarithmation, followed by a rescaling which broadens the distribution by a large factor, 
$E$ almost certainly takes a huge value, rather independently of the exact form of the distribution 
function $F_3$.

\section{Complying with Born's rule\label{Born}}

In section \ref{secRS} the wavefunction is modelled to split in two branches with equal 
amplitudes. Born's rule is trivially satisfied in this case. Does the model generalise correctly
to a splitting with unequal amplitudes?  
Technically, this is accomplished by
a unitary transformation devised in \cite{Zurek1998, Deutsch1999,Zurek2002} which entangles 
a two-state superposition with a large number of auxiliary states so as to form another equal-amplitude 
superposition. It will be argued that in this way the model scenario is consistent with Born's rule 
in general. 

The crucial point here is that ``if you believe in determinism, you have to believe it all 
the way'' \cite{tHooft2011}.
When an observer encounters a wavefunction for a measurement, like
\begin{equation} \label{SystemObserverBefore}
    a |A\rangle + b |B\rangle 
\end{equation} 
that wavefunction is given to him by the total operator of evolution. The operator is thus only required 
to handle wavefunctions that it provides itself. Extending the model accordingly would be based on the 
following considerations. 

In the course of measurement, a result $A$ or $B$ is obtained, but it always comes with many 
irrelevant properties of the constituents, like the number of photons scattered off the 
apparatus, the number of observer's neurons firing, etc. Let $n$ be the number of irrelevant 
properties to be taken into account. Then, after the measurement, we have a state vector of the 
form
\begin{equation}  \label{Result+Irrelevants}
    \sum_{k=1}^m c_k |A,k\rangle + \sum_{k=m+1}^n c_k |B,k\rangle
\end{equation}
The measuring evolution should commute with the projections on the spaces defined by $A$ and $B$, 
so we have constraints on the absolute values,
\begin{equation}  \label{ckConstraint}
   \sum_{k=1}^m |c_k|^2 = |a|^2  \qquad \qquad  \sum_{k=m+1}^n |c_k|^2 = |b|^2 
\end{equation} 
On the other hand, state vectors differing only in the phases of the $c_k$ can be regarded as
equivalent for the measurement process, as has been shown by different arguments in 
\cite{Zurek1998} and \cite{Deutsch1999}, and elaborately in \cite{Zurek2002}.   

Since the $k$-properties in (\ref{Result+Irrelevants}) are ``irrelevant'', we expect the
evolution to produce a state belonging to the equivalence class at the peak number of 
representatives. A measure of this number is given by the surface element in 
the space of $n$-dimensional normalised states  
$$
  \delta\left(1-\sum_{k=1}^n |c_k|^2\right) \prod_{k=1}^n \mathrm{d}^2 c_k
$$
which is defined uniquely, up to a constant, by its invariance under unitary changes of basis for 
the span of the vectors.  
The number of representatives is obtained by integrating over the phases, which gives
\begin{equation}  \label{ckSurface}
    \delta\left(1-\sum_{k=1}^n |c_k|^2\right) \prod_{k=1}^n 2\pi|c_k|\mathrm{d}|c_k|
\end{equation} 
At the maximum, all moduli must be nonzero because of the $|c_k|$ factors. It follows by the
permutation symmetry of constraints (\ref{ckConstraint}) that 
$$
   |c_k| = \frac{|a|}{\sqrt{m}} \quad k=1,\ldots,m \qquad
   |c_k| = \frac{|b|}{\sqrt{n-m}}\quad k=m+1,\ldots,n 
$$  
Finally extremising in the parameter $m$, by extremising the product of moduli in (\ref{ckSurface})
we find
\begin{equation}  \label{ckSpecified}
   |c_k| = \frac1{\sqrt{n}} ~~ \mbox{ for all }k 
\end{equation}
So the number $m$ of branches with property $A$ equals $|a|^2 n$. 
A similar argument, with a discussion of fluctuations about the maximum, was given for
a different scenario in \cite{Polley2005}.  
Equations (\ref{Result+Irrelevants}) and (\ref{ckSpecified}), in conjunction with an arbitrary 
choice of phases, like $c_k=\sqrt{1/n}$, now specify the state vector that an evolution operator for a
measurement should generate.   

Instead of splitting into ``up'' and ``down'' branches we now have splitting into $n$ 
branches, of which $m$ correspond to result $A$ of the measurement, and $n-m$ to result $B$. 
Information about $A$ or $B$ can be regarded as implicit in the indices of the records, so 
parameter $m$ need not even appear. 
In the equations of section \ref{secConstruction} the following replacements have to be made.
In (\ref{DefUO}), $\psi_u+\psi_d$ extends to a sum over $n$ equal parts. 
In (\ref{Register}) the entries $u,d$ of the register change to $1,\ldots,n$, and the same holds for 
the address sets (\ref{AkSuperposition}). 
Normalisation factors change from $\sqrt{1/2}$ to $\sqrt{1/n}$ in equations (\ref{DefUO}) 
and (\ref{UrwsPsi}).    
In section \ref{secExtremalBranch} the nonzero branching probabilities become $p_n \ll 1$
and $p_1 = 1-p_n$, and all powers of 2 change to powers of $n$.

\section{Conclusion\label{Conclusion}}

Observer's consciousness playing a role in the projection postulate 
has been pondered since the beginnings of quantum theory \cite{vNeumann1932}. 
The model presented here is a proposal of how the idea might be realised,
albeit with modifications, in a framework of unitary quantum-mechanical evolution.
 In the model scenario, the projection involved is 
on the conscious subspace at the time of the extremal draw as defined in section
\ref{secConsciousDimension}. But this projection assigns an outcome, ``up'' or ``down'', 
not only to a single measurement (or quantum event, as it was called here) but to all measurements 
of an observer's lifetime. Moreover, an assumption on statistics in the dynamics of 
consciousness was crucial to the functioning of the model. It might be objected then that we have only 
replaced the projection postulate with a statistical postulate based on speculation about consciousness. 
But there are well-known mechanisms for generating power-law statistics, some of which may be adaptable 
so as replace the postulate by the workings of an extended model. For the present model, it was important 
also to demonstrate that the preconditions of extreme-order theorems can be realised \emph{exactly} 
in a framework of unitary evolution. This was made obvious by employing random draws in the construction of 
the time-step operator.    
    
A rather complicated part in that construction, section \ref{RetrieveRecords}, 
was to identify by backward branching the records that had a chance to be written before a given record. 
This might suggest to alternatively \emph{accumulate} the extremal 
value in the course of evolution. But ``when the sum of [\ldots] independent heavy-tail random 
variables is large, then it is very likely that only one summand is large'' \cite{Barbe2004}, so the 
alternative approach is very likely to reduce to the one already taken.     

A scenario of consciousness, all generated by one extremal event in a short interval of time,
may also contribute to improving the physical notion of the metaphysical present. In the physical approach,
time is parameterised by the reading of a clock, and it is possible to quantify the time intervals of 
cognitive processing. But outside the science community, such an approach is often felt to inadequately 
represent the existential quality of a moment.  
The model scenario would suggest a more complex relation to physical time.
Phys\-ics\-wise, it is a moment indeed, but since it covers all individual experience, the moment
appears to be lasting.

The undifferentiated usage of ``consciousness'' in this paper will be unsatisfactory from a biological 
or psychological point of view, although consciousness as a processing of memories was discussed first in 
those fields \cite{Edelman1989}. For the present purpose, the 
meaning of the term was defined in section \ref{secConsciousDimension}.
As a consequence, activities of an observer which are conscious in the usual sense had to be regarded 
as ``unconscious''. Apparently, various levels of consciousness should be taken into account by 
an extended model. 

\newpage

\pdfbookmark{References}{}

\end{document}